\documentstyle[12pt,epsfig]{article}
\def\abstracts#1{{
	\centering{\begin{minipage}{12.2truecm}\footnotesize\baselineskip=12pt\noindent
	\centerline{\footnotesize ABSTRACT}\vspace*{0.3cm}
	\parindent=0pt #1
	\end{minipage}}\par}} 
 1
 1
 1

\newcommand{\nc}{\newcommand}
\nc{\renc}{\renewcommand}
\newlength{\undereqskip}
\nc{\be}[1]{\begin{equation} \mbox{$\label{#1}$}}
\nc{\bea}[1]{\begin{eqnarray} \mbox{$\label{#1}$}}
\nc{\eea}{\vspace{\undereqskip}\end{eqnarray}}
\nc{\ee}{\vspace{\undereqskip}\end{equation}}
\nc{\bc}{\begin{center}}
\nc{\ec}{\end{center}}
\nc{\ba}{\begin{array}}
\nc{\ea}{\end{array}}
\nc{\inv}[1]{\frac{1}{#1}}
\nc{\dbar}[2]{\frac{d\,^{#1}#2}{(2\pi)^{#1}}}
\def\Re{{\rm Re}\hskip2pt}
\def\Im{{\rm Im}\hskip2pt}
\def\simleq{\; \raise0.3ex\hbox{$<$\kern-0.75em
      \raise-1.1ex\hbox{$\sim$}}\; }
\def\simgeq{\; \raise0.3ex\hbox{$>$\kern-0.75em
      \raise-1.1ex\hbox{$\sim$}}\; }
\nc{\Tr}{{\rm Tr\,}}
\def\MeV{{\rm\ MeV}}
\def\half{\frac 1 2}
\def\ne{neutrino\ }
\def\nes{neutrinos\ }
\nc{\sign}{{\epsilon}}
\nc{\g}{\gamma}
\nc{\pub}[4]{\Bibitem{#1}#2, {\sl ``#3''}, #4.}
\nc{\advp}[3]{{\it  Adv.\ in\ Phys.\ }{{\bf #1} {(#2)} {#3}}}
\nc{\annp}[3]{{\it  Ann.\ Phys.\ (N.Y.)\ }{{\bf #1} {(#2)} {#3}}}
\nc{\apl}[3]{{\it  Appl. Phys. Lett. }{{\bf #1} {(#2)} {#3}}}
\nc{\apj}[3]{{\it  Astrophys.\ J.\ }{{\bf #1} {(#2)} {#3}}}
\nc{\apjl}[3]{{\it  Astrophys.\ J.\ Lett.\ }{{\bf #1} {(#2)} {#3}}}
\nc{\app}[3]{{\it Astropart.\ Phys.\ }{{\bf #1} {(#2)} {#3}}}

\nc{\cmp}[3]{{\it  Comm.\ Math.\ Phys.\ }{{ \bf #1} {(#2)} {#3}}}
\nc{\cqg}[3]{{\it  Class.\ Quant.\ Grav.\ }{{\bf #1} {(#2)} {#3}}}
\nc{\epl}[3]{{\it  Europhys.\ Lett.\ }{{\bf #1} {(#2)} {#3}}}
\nc{\ijmp}[3]{{\it Int.\ J.\ Mod.\ Phys.\ }{{\bf #1} {(#2)} {#3}}}
\nc{\ijtp}[3]{{\it Int.\ J.\ Theor.\ Phys.\ }{{\bf #1} {(#2)} {#3}}}
\nc{\jmp}[3]{{\it  J.\ Math.\ Phys.\ }{{ \bf #1} {(#2)} {#3}}}
\nc{\jpa}[3]{{\it  J.\ Phys.\ A\ }{{\bf #1} {(#2)} {#3}}}
\nc{\jpc}[3]{{\it  J.\ Phys.\ C\ }{{\bf #1} {(#2)} {#3}}}
\nc{\jpg}[3]{{\it J.~Phys.~G:~Nucl.~Part.~Phys.~}{{\bf #1} {(#2)} {#3}}}
\nc{\jap}[3]{{\it J.\ Appl.\ Phys.\ }{{\bf #1} {(#2)} {#3}}}
\nc{\jpsj}[3]{{\it J.\ Phys.\ Soc.\ Japan\ }{{\bf #1} {(#2)} {#3}}}
\nc{\lmp}[3]{{\it Lett.\ Math.\ Phys.\ }{{\bf #1} {(#2)} {#3}}}
\nc{\lncim}[3]{{\it  Lett.\ Nuov.\ Cim.\ }{{\bf #1} {(#2)} {#3}}}
\nc{\mpl}[3]{{\it  Mod.\ Phys.\ Lett.\ }{{\bf #1} {(#2)} {#3}}}
\nc{\ncim}[3]{{\it  Nuov.\ Cim.\ }{{\bf #1} {(#2)} {#3}}}
\nc{\np}[3]{{\it  Nucl.\ Phys.\ }{{\bf #1} {(#2)} {#3}}}
\nc{\pr}[3]{{\it Phys.\ Rev.\ }{{\bf #1} {(#2)} {#3}}}
\nc{\pra}[3]{{\it  Phys.\ Rev.\ }{{\bf A#1} {(#2)} {#3}}}
\nc{\prb}[3]{{\it  Phys.\ Rev.\ }{{\bf B#1} {(#2)} {#3}}}
\nc{\prc}[3]{{\it  Phys.\ Rev.\ }{{\bf C#1} {(#2)} {#3}}}
\nc{\prd}[3]{{\it  Phys.\ Rev.\ }{{\bf D#1} {(#2)} {#3}}}
\nc{\prl}[3]{{\it Phys.\ Rev.\ Lett.\ }{{\bf #1} {(#2)} {#3}}}
\nc{\pl}[3]{{\it  Phys.\ Lett.\ }{{\bf #1} {(#2)} {#3}}}
\nc{\prep}[3]{{\it Phys\. Rep.\ }{{\bf #1} {(#2)} {#3}}}
\nc{\prsl}[3]{{\it Proc.\ R.\ Soc.\ London\ }{{\bf #1} {(#2)} {#3}}}
\nc{\ptp}[3]{{\it  Prog.\ Theor.\ Phys.\ }{{\bf #1} {(#2)} {#3}}}
\nc{\ptps}[3]{{\it  Prog\ Theor.\ Phys.\ suppl.\ }{{\bf #1} {(#2)} {#3}}}
\nc{\physa}[3]{{\it  Physica\ A\ }{{\bf #1} {(#2)} {#3}}}
\nc{\physb}[3]{{\it  Physica\ B\ }{{\bf #1} {(#2)} {#3}}}
\nc{\phys}[3]{{\it Physica\ }{{\bf #1} {(#2)} {#3}}}
\nc{\rmp}[3]{{\it  Rev.\ Mod.\ Phys.\ }{{\bf #1} {(#2)} {#3}}}
\nc{\rpp}[3]{{\it Rep.\ Prog.\ Phys.\ }{{\bf #1} {(#2)} {#3}}}
\nc{\sjnp}[3]{{\it Sov.\ J.\ Nucl.\ Phys.\ }{{\bf #1} {(#2)} {#3}}}
\nc{\spjetp}[3]{{\it Sov.\ Phys.\ JETP\ }{{\bf #1} {(#2)} {#3}}}
\nc{\yf}[3]{{\it Yad.\ Fiz.\ }{{\bf #1} {(#2)} {#3}}}
\nc{\zetp}[3]{{\it Zh.\ Eksp.\ Teor.\ Fiz.\ }{{\bf #1} {(#2)} {#3}}}
\nc{\zp}[3]{{\it Z.\ Phys.\ }{{\bf #1} {(#2)} {#3}}}
\nc{\zpc}[3]{{\it Z.\ Phys.\ C\ }{{\bf #1} {(#2)} {#3}}}
\nc{\ibid}[3]{{\sl ibid.\ }{{\bf #1} {#2} {#3}}}
\begin{document}
\rightline{FTUV/98-58; IFIC/98-59}
\vskip 0.7cm
\centerline{\normalsize\bf RADIATIVE NEUTRINO DECAY IN MEDIA}
\baselineskip=16pt
\vspace*{0.6cm}
\centerline{Dario Grasso \footnote{New address since 1st October 1998:
Dipartimento di Fisica, Universit\`a di Padova, Via F. Marzolo 8, I-35131
Padova, Italy}}
\baselineskip=13pt
\centerline{\footnotesize\it Departament de Fisica Teorica, 
Universitat de Valencia}
\baselineskip=12pt
\centerline{\footnotesize\it E-46100 Burjassot - Valencia, Spain}
\centerline{\footnotesize E-mail: grasso@flamenco.ific.uv.es}
\vspace*{0.3cm}
\centerline{\footnotesize and}
\vspace*{0.3cm}
\centerline{Victor Semikoz}
\baselineskip=13pt
\centerline{\footnotesize\it Institute of Terrestrial Magnetism, the 
Ionosfere and Radio Wave Propagation}
\baselineskip=12pt
\centerline{\footnotesize\it Academy of Science of Russia, Troitsk, Moscow 
Region, 142092 Russia}
\centerline{\footnotesize E-mail: semikoz@izmiran.rssi.ru}
\vskip 1.cm
\abstracts{\footnotesize
In this letter we introduce a new method to determine the 
radiative neutrino decay rate in the presence of a medium. 
Our approach is based on the generalisation of the optical theorem at finite 
temperature and density.
Differently from previous works on this subject, our method allows to
account for dispersive and dissipative electromagnetic properties of the medium.
Some inconsistencies that are present in the literature are pointed-out
and corrected here.
We shortly discuss the relevance of our results for neutrino evolution in the 
early universe.}
\normalsize\baselineskip=15pt
\section{Introduction}
Most of \nes in the universe are expected to be produced in the 
presence of environments characterised by very high temperatures and 
densities. In such media \nes behave quite differently from the 
way they do in vacuum \cite{raffeltbook}. The coherent interaction of \nes with
the particles belonging to the medium can amplify or induce effects 
that are otherwise very feeble or absent in the vacuum.
This can be crucial in order to disentangle possible anomalous properties
of the neutrinos that may testify for new physics present beyond the standard 
model. MSW \cite{MSW} \ne resonant oscillations give a well know example 
of such a  kind of effects.  

Neutrino electromagnetic properties are also expected to be significantly
affected by the presence of media.  In fact, due to their coherent interaction
with charged leptons and nucleons in the medium, \nes acquire an effective 
coupling to the electromagnetic field.  This effect was investigated 
by several authors who showed how even massless neutrinos passing through 
the matter acquire an effective charge 
\cite{OraSem87} and, in a charge asymmetric medium, 
also an effective magnetic dipole moment \cite{Victor87}.
Several intriguing consequences of medium induced \ne electromagnetic 
couplings have been studied in the literature, {\it e.g.} 
plasmon decay in $\nu{\bar \nu}$ pairs \cite{pldecay} and Cherenkov emission 
of \nes \cite{Cherenkov}.

In this letter we focus on the radiative decay of a heavy \ne into a 
lighter one. Our underling particle physics model consists of  
a minimal extension of the standard model which allows for non-vanishing \ne 
masses and mixing.
In the ambit of this model, \ne radiative decay is extremely suppressed in 
vacuum because of the Glashow-Iliopoulos-Maiani (GIM) cancellation 
\cite{PalWol}. However, if $T < m_{\tau}$, this is not the case
in a heat-bath due to the different thermal populations of the three 
charged leptons families.
As a consequence, the radiative \ne decay rate in the presence of hot media 
exceeds the decay rate in vacuum by many order of magnitudes.  
This was first showed by D'Olivo, Nieves and Pal (DNP) \cite{DNP}
(see also \cite{GiuntiKL}), 
who computed the decay rate both in the case of a non-relativistic plasma 
(NR) ($T \ll m_e$) and of a ultra-relativist plasma (UR).
Recently, DNP's work has been extended by
Nieves and Pal \cite{NiePal} (NP) who showed that the decay rate may be 
further enhanced due to the Bose-Einstein stimulation in the production of
low-energy photons ($\omega \ll T$).  

Although we agree with DNP and NP concerning the huge neutrino decay rate
amplification occuring in a medium we do not , however, about their
method and some of the results obtained in their works. 
Our main disagreement concerns the properties of the photon produced 
by the \ne decay. This photon was assumed
to be on the light-cone ($K^2 = 0$) by DNP and NP. 
Such an assumption is equivalent to
disregard any dispersive and dissipative property of photons in a 
thermal bath. 
It is  well known, however, that in plasma photons behaves more like 
collective excitations (usually dubbed {\it plasmons}) than like 
free particles.
While neglecting collective properties of photons 
may be a reasonable approximation in the high frequency regime this is
however inconsistent when they are low energy or, so-called, soft. 
We will show that to account for the correct plasmon dispersion relation
is indeed mandatory in order to avoid an unphysical behaviour of 
the decay rate as a function of the \ne velocity. 
 
Photon dissipative properties need also to be taken into account.  
In fact, in a heat bath low-energy  photons 
are continuously absorbed and re-emitted mainly by Landau-damping on the
heat-bath free-charges. 
For the first time we will show that Landau-damping gives rise to an
additional, and in many cases dominant, contribution to the decay rate.
Since the photon asymptotic state is a not a well defined quantity in a medium,
we recourse here to a treatment that is based on finite temperature 
field theory and on the generalisation of the optical theorem to such 
a framework \cite{Weldon83}. Within this framework the basic quantities are the 
Green functions which are well defined quantities even at finite temperature 
and density.
As a bonus, one get that using such approach the role of Landau-damping 
comes-out naturally from the thermally-corrected photon progator 
\cite{Altherr91}. 
A similar method has been previously applied by other authors to determine the 
chirality-flip rate of Dirac \nes in a degenerate plasma \cite{AltKai,AyaDT} 
and in the early universe \cite{ElmERS}.

In this letter we only consider the case of an UR non-degenerate 
electron-positron plasma. As we will shortly discuss in our last 
section, this case is relevant for the study of neutrino evolution in the 
early universe. The cases in which the medium  
is degenerate and/or NR will be discussed elsewhere \cite{GraSem}. 
Our general method will be presented in section 2. Section 3 contains
the computation of the decay rate in the case in which the plasmon is on-shell.
The generalisation of this result to the case with off-shell photons is
discussed in Sec. 4. Sec. 5 contains a discussion about the 
applications of our results to the early universe physics. 

\section{The general method}
It was first showed by Weldon \cite{Weldon83} that  
in a heat-bath the decay rate $\Gamma_d$ of a fermionic particle 
species (having the energy $E$), 
and the rate of the inverse process $\Gamma_i$, are related to the imaginary 
part of the fermion self-energy $\Sigma$ by the following relation 
\be{weldon}
\Im \left\{ {\overline{u}_(p)}\Sigma(p) u(p) \right\} =
- E(\Gamma_i + \Gamma_d)~.
\ee
This result is the generalisation of the optical theorem for finite 
temperature and density. It is understood that $\Sigma$ includes 
corrections due to the effects of the medium on the field propagators. 

In our case, the interesting contribution to the heavy neutrino self-energy
comes from the Feynman diagram represented in Fig. 1. In that figure $\nu_i$ 
and $\nu_j$ are, respectively, the heavy and the light neutrino mass eigenstates
participating to the decay and inverse-decay processes. 
Working in the real-time-formalism (RTF) we determine the imaginary part
of $\Sigma$ by properly cutting \cite{KobesS85} the diagram represented 
in Fig. 1.
\begin{figure}
\begin{center}
\vspace*{13pt}
\epsfig{file=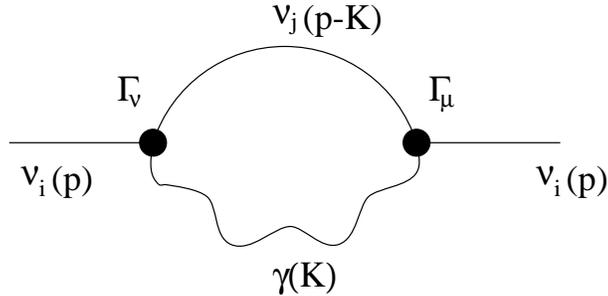,width=8cm}
\caption{Feynman diagram contributing to the heavy neutrino self-energy 
$\Sigma$. }
\label{fig:1}
\end{center}
\end{figure}
We get
\be{sigma}  
\Im \Sigma(p) = - \frac{\sign(p_0)}{2 \sin 2\phi_p} 
\int \dbar{4}{K} \Gamma^{\nu\,*}(K) \left[ S^-(p - K)D^+_{\mu\nu}(K)
+ S^+(p - K)D^-_{\mu\nu}(K) \right] \Gamma_\mu(K)
\ee
where 
\be{sindef}
   \inv{2}\sin 2\phi_p=\frac{e^{\beta|p_0|/2}}{e^{\beta|p_0|}+1}
\ee
and the expression for the effective vertex  $\Gamma_\mu(K)$ 
will be given below. Here $K = (k_0, {\vec k})$ is the photon 
4-momentum in the medium rest-frame and $\sign(p_0)$ is the sign function.
We observe that a positive $k_0$ accounts for the process with
emission of a photon (energy given to the medium) whereas a negative
$k_0$ takes care of the inverse process (energy lost by the medium).
We are then able to disentangle the different contributions to $\Gamma_d$ and
$\Gamma_i$ from (\ref{weldon}) by properly choosing the $k_0$ 
integration interval.
For the sake of simplicity we assume the light neutrino
to be massless. In the RTF the off-diagonals component of its propagator are 
\be{Spm}
S^\pm(p - K) = -2\pi i\, ({\hat p} - {\hat K})\, 
\left( \theta(\pm (p_0 - k_0) - n_F(|p_0 - k_0|)\right) 
\delta\left( (p - K)^2\right)
\ee
where $n_{F (B)}$ is the Fermi-Dirac (Bose-Einstein) distribution function.
In the rest frame of the medium (and in the Landau gauge)
the photon propagator can be decomposed into 
\be{Dpm1}
D^\pm_{\mu\nu}(K) = D^\pm_T P_{\mu\nu}(K) + D^\pm_L Q_{\mu\nu}(K)~.
\ee
The transverse and longitudinal projector operators are defined by
\be{proj} 
P_{\mu\nu}(K) = \left(\delta_{ij} -  \frac{k_i k_j}{k^2} \right) 
\delta_{\mu i}\delta_{\nu j} \qquad 
Q_{\mu\nu}(K) =  e_\mu(K) e_\nu(K)
\ee
where $e_\mu(K) = (k,\ k_0{\vec k}/k)/\sqrt{K^2}$ is the polarization versor.
The coefficients $D^\pm_{T,L}$ are given by \cite{Altherr91,Lemoine}
\be{Dpm2}
\ba{rcl}\displaystyle
D^\pm_{T,L} = 2  \left( \theta(\pm k_0 - n_B(|k_0|) \right)
\Big[&& \pi \delta\left(K^2 - \Re\Pi_{T,L}(K)\right) 
\delta_{Kr}(\Im\Pi_{T,L}(K))  \\[5mm]
 &-& \displaystyle \frac{\Im\Pi_{T,L}(K)}{ \left(K^2 - 
\Re\Pi_{T,L}(K)\right)^2 +
\left(\Im\Pi_{T,L}(K)\right)^2 } \Big]
\ea
\ee
where $\delta_{Kr}$ is a Kroneker function. 
These expressions include the one-loop thermal corrections to the photon 
propagator. It is known \cite{BraatenP90} that for soft photons 
($k_0, k \simleq eT$) these corrections cannot be neglected  
and that a meaningful perturbative expansion can be obtained performing a  
proper resummation of the hard-thermal-loops contributing to the photon 
polarization tensor. 

Equation (\ref{Dpm2}) has been written in such a way in order to distinguish
an {\it on-shell} part, corresponding to the propagation of the 
transversal and longitudinal modes of a plasmon, and an {\it off-shell}
part associated to Landau-damping. 
The fluctuation-dissipation theorem  shed light on
the physical nature of off-shell plasmons: they are 
thermal fluctuations of the electromagnetic field \cite{Lemoine}. 
As suggested by the Breit-Wigner form of the off-shell part of 
$D^\pm_{T,L}$, we can think of electromagnetic thermal fluctuations as 
resonances which are continuously emitted 
and re-absorbed from the free-charges in the plasma by Cherenkov and 
inverse-Cherenkov emission.
  
The general expression of the polarization tensor in a isotropic plasma
is
\be{Pi}
\Pi_{\mu\nu}(K) = \Pi_T(K) P_{\mu\nu}(K) + \Pi_L(K) Q_{\mu\nu}(K)
\ee
where, for an UR plasma, we have \cite{Weldon82} 
\be{PiTL}
\ba{rcl}\displaystyle
   \Re\Pi_T(K)&=&\displaystyle\frac{3 \omega_P^2}{2}
        \left[\frac{k_0^2}{k^2}+(1-\frac{k_0^2}{k^2})
        \frac{k_0}{2k} \ln\left|\frac{k_0+k}{k_0-k}\right|\right]~~, \\[5mm]
   \Re\Pi_L(K)&=&\displaystyle 3\omega_P^2
        (1-\frac{k_0^2}{k^2})
        \left[1-\frac{k_0}{2k}
        \ln\left|\frac{k_0+k}{k_0-k}\right|\right]
\ea
\ee
and
\be{PiTLim}
\ba{rcl}\displaystyle
   {\rm Im}\,\Pi_T(K)&=&\displaystyle-\frac{3\omega_P^2}{2}\pi
        (1-\frac{k_0^2}{k^2})\,\frac{k_0}{2k}\,\theta(k^2-k_0^2)~~, \\[5mm]
   {\rm Im}\,\Pi_L(K)&=&\displaystyle 3\omega_P^2\pi
        (1-\frac{k_0^2}{k^2})\,\frac{k_0}{2k}\,\theta(k^2-k_0^2)~~.
\ea
\ee
In the UR limit, the plasma frequency is given by $\omega_P = e T/3$.

Plasmon dispersion relations are determined by solving $K^2 = \Re\Pi_{T,L}(K)$.
Analytic forms of the solutions can be obtained in some suitable limits
\cite{Altherr91}.
In the $k \rightarrow 0$ limit transversal and longitudinal dispersion
relations are respectively
\be{dispsoft}
\ba{rcl}
   k_0^2 &=& \omega_P^2 \displaystyle+\frac{6}{5}k^2 \qquad 
{\rm transversal}~~, \\[5mm]
k_0^2 &=& \omega_P^2 \displaystyle+\frac{3}{5}k^2 \qquad
{\rm longitudinal}~~,
\ea
\ee
whereas in the hard limit ($k_0, k \gg \omega_P$) one gets
\be{disphard}
\ba{rcl}\displaystyle
   k_0^2 &=& \displaystyle \frac 3 2 \omega_P^2 + k^2 \qquad 
~~~~~~~~~~ {\rm transversal}~~, \\[5mm]
k_0 &=& k^2  + 4 k^2 e^{-2 k^2/3\omega_P} \qquad
{\rm longitudinal}~~.
\ea
\ee
 
\section{The on-shell plasmon case}
We now come into some more details of our computation.
As we discussed in the introduction, in a medium the
neutrino-photon coupling is mediated by the weak interaction of the
neutrino to the free charges in the plasma. We the sake of simplicity we
only consider here a non-degenerate electron-positron plasma at temperatures 
below the muon mass.
From \cite{OraSem87,EnqKS} we know that the effective medium induced neutrino 
electromagnetic vertex is given by 
\be{vertex} 
\Gamma_{\mu}(K) =  U_{ie} \frac{G_F}{e} \Pi_{\mu\nu}(K) \g^\nu L
\ee 
where $U$ is the lepton mixing matrix and $L \equiv \half (1 - \g_5)$.

By substituting (\ref{vertex}), (\ref{Spm}) and the {\it on-shell} part
of (\ref{Dpm2}) in (\ref{sigma}), we find that the decay rate of the 
heavy neutrino into the lighter one and a transversal plasmon is
\be{gammaT} 
\ba{rcl}\displaystyle
\Gamma_{T}^{on} = \frac{1}{2\sin2\phi_p}&& \displaystyle \frac{ G_F^2}{4E e^2}  
\vert U^*_{ie}U_{je} \vert^2  \int \dbar{4}{K} (\Re\Pi_T(K))^2
\delta\left(K^2 -  \Re\Pi_T(K)\right) \delta\left( (p - K)^2 \right)\\[5mm]
&& \times \left[ \sign(p_0 - k_0) (\theta(- k_0)+ n_F(|p_0 - k_0|)) 
+ \sign(p_0 - k_0)(\theta(k_0) +  n_B(|k_0|)) \right] \\[5mm] 
&& \times P_{\alpha\beta} \Tr \left\{({\hat p} - m)\g^\beta({\hat p} - 
{\hat K})\g^\alpha L \right\}~~
\ea
\ee
where $m$ is the heavy neutrino mass. 
In the case the plasmon is longitudinal the rate can be obtained by replacing 
in (\ref{gammaT}) the subscript $T$ with $L$ and the projector 
$P_{\alpha\beta}$ with $Q_{\alpha\beta}$.
In the square brackets in the right side of (\ref{gammaT}) 
we can distinguish a term accounting for the Bose-Einstein stimulated 
photon emission and a Pauli-blocking term for the production of the 
light \ne. Since generally the latter is much smaller than the former
we disregard the Pauli-blocking term in the following.
We also approximate the factor $\displaystyle \frac{1}{\sin2\phi_p}$ with 
the unity.

Using $\delta\left( (p - K)^2 \right)$ and 
$\delta\left(K^2 -  \Re\Pi_T(K)\right)$ to suppress respectively the angular 
and the temporal parts of the $\displaystyle \dbar{4}{K}$ integration we get
\be{rate-on}
\Gamma_{T}^{on} = \frac{9 G_F^2}{32 \pi e^2} \vert U^*_{ie}U_{je} \vert^2
\omega_P^5  f_{T}(v)~~.
\ee
Here $v$ is the velocity of the heavy neutrino with respect to the medium
rest frame.
Since it is in general impossible to find an analytical form of the
plasmon dispersion relation, it is convenient 
to distinguish between a soft and a hard contribution to (\ref{rate-on}).
In the case the plasmon is transversal we have
\be{fvT} 
 f_{T}(v) =  f^{\rm hard}_{T}(v) + f^{\rm soft}_{T}(v)
\ee
\be{fvtsoft}
f^{\rm soft}_{T}(v) \simeq \frac 8 9 \frac{1 - v^2}{{\tilde m^2} v} 
\int_{{\tilde k}^{\rm soft}_{min}}^{{\tilde k}^{\rm soft}_{max}}
d{\tilde k} (1 + n_B(1)) \left\{{\tilde E}^2 - \half (1 + {\tilde m}^2) -
 \frac{1}{{\tilde k}^2}  \left[{\tilde E} -
\half(1 + {\tilde m}^2) \right] \right\} \theta(1 - {\tilde k})
\ee
and
\be{fvthard}
f^{\rm hard}_{T}(v) \simeq \frac{1 - v^2}{{\tilde m^2} v} 
\int_{{\tilde k}^{\rm hard}_{min}}^{{\tilde k}^{\rm hard}_{max}}
\frac{d{\tilde k}}{{\tilde k}} (1 + n_B({\tilde k})) \left\{
{\tilde E} - \frac {{\tilde k}^2}{2}  -
 \frac{{\tilde m}^2}{4{\tilde k}} \right\} \theta({\tilde k} - 1)
\ee
where $E$ is the energy of the heavy neutrino and
tilted quantities have been normalised to $\omega_P$ so that 
$f_{T}(v)$ is a dimensionless function.
The integration limits are
\be{klimhard}
{\tilde k}^{\rm hard}_{max} = \frac{\tilde m}{2}\sqrt{\frac{1 + v^2}{1 - v^2}} 
\qquad
{\tilde k}^{\rm hard}_{min} = \frac{\tilde m}{2}\sqrt{\frac{1 - v^2}{1 + v^2}} 
~~,
\ee
in the hard limit, and
\be{klimsoft}
{\tilde k}^{\rm soft}_{max} = {\tilde m}\sqrt{\frac{1 + v^2}{1 - v^2}} - 1
\qquad
{\tilde k}^{\rm soft}_{min} = \left| {\tilde m}\sqrt{\frac{1 - v^2}{1 + v^2}} 
- 1 \right|~~,
\ee
in the soft one.
For reason of available space, only a zero order expansion in ${\tilde k}$
of $\Re\Pi_T(K)$ has been reported in the expressions (\ref{fvthard}) and 
(\ref{fvtsoft}). However, a more complete second order expansion has been
used to get Fig. 2.
\begin{figure}
\begin{center}
\vspace*{13pt}
\epsfig{file=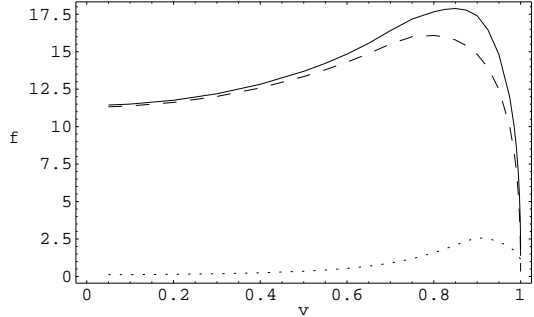,width=7cm}
\caption{In this figure the dashed, dotted and continuous lines represents 
respectively the functions $f^{\rm hard}_{T}(v)$, $f^{\rm soft}_{T}(v)$ 
and the sum of the two. Here we fixed ${\tilde m} = 5$.}
\label{fig:2}
\end{center}
\end{figure}
Also, in our numerical integrations some more suitable exponential cut-off 
functions have been used instead of the step functions appearing in the 
integrals in (\ref{fvthard}) and (\ref{fvtsoft}).
From Fig. 2 we see that in spite of the large enhancement of the Bose-Einstein 
factor  $n_B(|k_0|) = (e^{\beta|k_0|}-1)^{-1} \simeq T/k_0 \gg 1$ that one 
gains in the soft limit, the soft contribution to the total decay 
rate is subdominant with respect to the hard one. This is due to the smaller 
available phase space when the plasmon is  soft.
Not surprisily, such a situation becomes even more
pronounced for higher values of the decaying neutrino mass.
The maximum value of $f^{\rm soft}_{T}$ is reached at the velocity
$\displaystyle v_* \equiv \frac{m^2 - \omega_P^2}{m^2 + \omega_P^2}$
where ${\tilde k}^{\rm soft}_{min} = 0$. 

It is worthwhile to observe that, a part for the presence of the infra-red 
cut-off function  (and of an extra factor of 2), our expression 
(\ref{fvthard}) coincides with the result found by NP. 
In our opinion, the introduction of the cut-off is essential in
order to avoid the incorrect use of the hard plasmon dispersion relation 
($K^2 \simeq 0$) in the soft limit.
In fact, in the limit in which the decaying neutrino is  ultra-relativistic 
($v \rightarrow 1$) it is evident from (\ref{klimhard}) that 
${\tilde k}^{\rm hard}_{min} \rightarrow 0$. 
As we have seen above, our approach solves such a problem by properly separating 
the hard and the soft contributions to $f_T(v)$.
To confirm us of the validity of our method, we get a physically
correct behaviour of the decay rate that drops to zero when 
$v \rightarrow 1$  (see Fig. 2). Such a result is actually convincing since 
no rest frame is 
available for the neutrino to decay in when it travels at the speed of light. 
A non-vanishing rate was instead found by NP in the limit $v \rightarrow 1$.

In the case the plasmon is longitudinal
only the soft term contributes to the total decay rate.
In fact, as we can see from the second of the equations (\ref{disphard})     
$K^2$, hence also $\Re\Pi_L(K)$, vanishes in the hard limit.
Due to angular momentum conservation,  $\Gamma_L(K)$ is suppressed by
a factor $(\omega_P/m)^2$ with respect to the soft contribution to
$\Gamma_T(K)$. For this reason its contribution to
the total decay rate is  subdominant. 
For reason of available space the expression of
$\Gamma_L(K)$ will be presented elsewhere \cite{GraSem}.

\section{The off-shell plasmon case}
In this section we discuss the contribution of thermal fluctuations of the
electromagnetic field to the neutrino decay. As we discussed in the Sec. 2  
the effect of these fluctuations is accounted by the second term  
on the right side of (\ref{Dpm2}).  
Differently from the case considered in the previous section, 
plasmons associated to thermal fluctuations are not {\it on-shell}
hence they do not obey any dispersion relation.
As a consequence, in these case we have to deal with a double 
integration both over $k$ and $k_0$.
Indeed, the expressions for the decay rate look in this case
\be{rate-off}
\Gamma_{T,L}^{off} = \frac{9 G_F^2}{32 \pi e^2} \vert U^*_{ie}U_{je} \vert^2
\omega_P^5  g_{T,L}(v)~~.
\ee     
The expression of the kinematical functions $g_{T,L}(v)$ are
\be{gvt}
\ba{rcl}\displaystyle
g_{T}(v) \simeq 
2{\tilde T} \frac{1 - v^2}{{\tilde m^2} v}
\int_0^\infty d{\tilde k}_0  
\int_{{\tilde k}^{\rm soft}_{min}}^{{\tilde k}^{\rm soft}_{max}}
\frac {d{\tilde k}}{{\tilde k}^2} && \displaystyle \left[ {\tilde E}^2 -
\frac {({\tilde K}^2 + {\tilde m}^2)}{2} - \frac{\left({\tilde E}{\tilde k}_0 
- \half ({\tilde K}^2 + {\tilde m}^2)\right)^2}{{\tilde k}^2} \right]\\[5 mm] 
&\times&{\tilde K}^2  A_T({\tilde k}_0,
{\tilde k}) \theta({\tilde k}^2 - {\tilde k}_0^2)
\ea
\ee
and 
\be{gvl}
\ba{rcl}\displaystyle
g_{L}(v) \simeq \frac {\tilde T}{2} \frac{1 - v^2}{{\tilde m^2} v}
\int_0^\infty d{\tilde k}_0  
\int_{{\tilde k}^{\rm soft}_{min}}^{{\tilde k}^{\rm soft}_{max}}
\frac {d{\tilde k}}{{\tilde k}^4} && \displaystyle \left[ \left(2{\tilde E} -
{\tilde k}_0\left(1 + \frac{{\tilde m}^2}{{\tilde K}^2}\right)\right)^2 -
\left({\tilde K}^2 - {\tilde m}^2\right){\tilde k}^2 \right] \\[5 mm]
&\times&{\tilde K}^4  A_L({\tilde k}_0,
{\tilde k}) \theta({\tilde k}^2 - {\tilde k}_0^2)
\ea
\ee
where
\be{ATL}
A_{T,L}({\tilde k}_0,{\tilde k}) \equiv
\frac{\left(\Re\Pi_{T,L}(K)\right)^2 + \left(\Im\Pi_{T,L}(K)\right)^2 }
{ \left(K^2 - \Re\Pi_{T,L}(K)\right)^2 +
\left(\Im\Pi_{T,L}(K)\right)^2 }~~. 
\ee
Since Landau-damping is effective only for soft photons we can 
neglect here any contribution coming from the hard part of the photon spectrum.
This allowed us to perform the substitution 
$(1 + n_B(k_0)) \rightarrow T/k_0$ in (\ref{gvt},\ref{gvl}).
The functions $g_{T,L}(v)$ have been computed numerically and their
plots are reported in Fig. 3. From this figure  
we see that like for the soft contribution to $\Gamma_{T,L}^{on}$,
$\Gamma_{T,L}^{off}$ reach their common maximum value
when $v = v_*$. This is not unexpected
since the functions $A_{T,L}(K)$ take their maximum values when $K^2$ 
approaches $\Re\Pi_{T,L}(K)$.
\begin{figure}
\begin{center}
\vspace*{13pt}
\epsfig{file=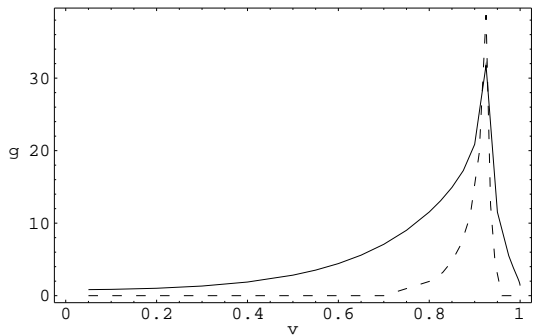,width=7cm}
\caption{In this figure the dashed and the continuous lines represents 
respectively the functions $g_{T}(v)$, $g_{L}(v)$. 
 Again, we fixed here ${\tilde m} = 5$. }
\label{fig:3}
\end{center}
\end{figure}

We have, however, a crucial difference between the off-shell and
the on-shell contributions to the total decay rate. 
Whereas radiative \ne decay is kinematically forbidden if 
$K^2 > m^2$  this is clearly not the case whenever the photon is off-shell.
Note that even in the hard limit, where $K^2 \simeq \displaystyle \frac 3 2
\omega_P^2$ (see (\ref{disphard})), 
the {\it on-shell} decay cannot take place unless 
$m^2 > \displaystyle \frac 3 2 \omega_P^2$. 
This was not noted in \cite{DNP,GiuntiKL,NiePal}.
We have to mention, however, that although $\Gamma^{off}$ remains different 
from zero it is typically very small when $m \ll \omega_P$. 

\section{Discussion}
The results of our previous sections can be directly applied to study
neutrino evolution in the early universe where, for $T \gg 1$ \MeV, the
electron-positron plasma is UR and non-degenerate. 
For definitess we consider the radiative decay of a tau-neutrino with
a mass in the \MeV\  range into a massless $\nu_e$.
In order to determine if radiative decay can give rise to any relevant 
depletion of tau-neutrinos we have to compare the decay rate
with the expansion rate of the universe.
Using our previous results we find that the decay rate is given by  
\be{finalrate}
\Gamma \simeq 2\times 10^{-5} \left(\frac {T}{1 \MeV}\right)^{5}
\vert \sin^{2}2\theta \vert \frac {f(v) + g(v)}{10}~~s^{-1} 
\ee
where $\theta$ is the vacuum mixing angle between the $\nu_e$ and the 
$\nu_\tau$ and $f(v)$ and $g(v)$ are the sum of the transversal and
longitudinal kinematical functions computed in the previous sections.
As we showed above $f(v) + g(v)$
is typically of the order of few tens if $m \approx T$.  
At the big-bang nucleosynthesis
(BBN) time the Hubble rate is given by
\be{hubble}
H \simeq 1 \left(\frac {T}{1 \MeV}\right)^{2}~~s^{-1}.
\ee
It is easy to verify that $\Gamma$ can exceed $H$ only for temperatures
that are above (though not too much) the range at which BBN take place
($0.1 \simleq T_{BBN} \simleq 10$ \MeV). 
Furthermore, even if $\Gamma > H$ at such high temperatures, one should not 
forget that no \ne depletion could take place since, for  $T \simgeq 2 m$,  
standard weak processes are always able to re-produce tau-neutrinos. 

A different scenario might be realised in the case one considers a larger 
extension of the standard model which allows for the presence of a neutrino
transitional magnetic moment. A magnetic moment induced decay can
become dominant with respect to that induced by the medium when the plasma
in NR and neutrinos are out of thermal equilibrium. 
Although Landau-damping gives an 
extra contribution to the decay also in this case, we do not expect that this 
effect could give rise to a significant neutrino depletion after BBN.
We observe, however, that astrophysical and cosmological constraints on the \ne 
transitional magnetic moments, that have been previously derived disregarding 
dispersive and dissipative properties of the medium, should be reconsidered in 
the light of the results obtained in this work.
This, as other related issues, will be discussed in a forthcoming 
paper \cite{GraSem}.

\section*{Acknowledgements}
The work of D.G. have been supported by the TMR network grant 
ERBFMRXCT960090 of the European Union, and that of V.S. by INTAS grant
96-0659 and by the sabbatical grant SAB95-506 and RFFR 97-02-16501, 
95-02-03724.
\newpage
\noindent

\end{document}